\documentstyle[prb,aps,epsf]{revtex}

\begin{document}

\twocolumn[
\hsize\textwidth\columnwidth\hsize\csname@twocolumnfalse\endcsname

\title{Continuous measurements of two qubits}

\author{Wenjin Mao and Dmitri V. Averin}

\address{Department of Physics and Astronomy, Stony Brook
University, SUNY, Stony Brook, NY 11794-3800}

\author{Francesco Plastina}

\address{Dipartimento di Fisica, Universita' della Calabria,
I-87036 Arcavacata di Rende (CS), Italy}

\author{Rosario Fazio}

\address{NEST-INFM $\&$ Scuola Normale Superiore, 56126 Pisa,
    Italy}

\date{\today}

\maketitle

\begin{abstract}
We develop a theory of coherent quantum oscillations in two, in
general interacting, qubits measured continuously by a mesoscopic
detector with arbitrary non-linearity and discuss an example of
SQUID magnetometer that can operate as such a detector. Calculated
spectra of the detector output show that the detector non-linearity
should lead to mixing of the oscillations of the two qubits. For
non-interacting qubits oscillating with frequencies $\Omega_1$ and
$\Omega_2$, the mixing manifests itself as spectral peaks at the
combination frequencies $\Omega_1\pm \Omega_2$. Additional
nonlinearity introduced by the qubit-qubit interaction shifts all
the frequencies. In particular, for identical qubits, the
interaction splits coherent superposition of the single-qubit
peaks at $\Omega_1=\Omega_2$. Quantum mechanics of the measurement
imposes limitations on the height of the spectral peaks.
\end{abstract}

\pacs{PACS numbers: 03.65.Ta; 03.67.Lx; 73.23.-b}

]

\section{Introduction}

Quantum measurements represent an important part of quantum
information processing, quantum computing, and evolution of
quantum systems in general. Counter-intuitive features of
the measurement process related to the collapse of the
wave-function of the measured system continue to attract
interest to the ``problem of quantum measurements''. Physics of
mesoscopic solid-state qubits and detectors provides convenient
tools for studying this problem in the most interesting case
of quantum systems that are large on atomic scale -- see, e.g.,
chapters on quantum measurements in \cite{b1}. Many features of
quantum measurements manifest themselves directly in the regime
of ``continuous'' measurements in which the measured system
evolves in time being continuously affected by the detector
back-action which implements the wave-function collapse
dynamically. Simple example of this regime is provided by the
linear weak measurements of coherent quantum oscillations in
one qubit \cite{b3,b4,b11,b5,b7} which have been demonstrated
experimentally in \cite{b6}. An interesting property of the
continuous weak measurements is that the spectrum of the
detector output has features that characterize directly the
quantum mechanics of measurement and quantum nature of the
qubit oscillations.

One of the suggested tools in quantum-information applications
of quantum measurements are quadratic measurements, realized
when the detector response to the input signal is quadratic.
Quadratic measurements should make it possible to monitor
products of operators of different quantum systems, and
can be used, for instance, to implement simple schemes of
error-correction \cite{err}, or to entangle non-interacting
qubits \cite{b12}. The purpose of our work is to study the
regime of continuous quadratic measurements concentrating on
the case of two qubits which is the simplest system that
reveals non-trivial characteristics of quadratic measurements.
Currently, systems of two coupled mesoscopic qubits are almost
routinely studied in experiments -- see
\cite{b14,b15,b16,b17,b18}. As discussed in this
work, for two qubits, quadratic measurements are equivalent
to measurements with arbitrary nonlinearity. Qualitatively,
the main feature of continuous non-linear measurements is the
mixing of coherent oscillations in individual qubits that
for non-interacting qubits leads to appearance of spectral
peaks at combination frequencies $\Omega_1\pm \Omega_2$, where
$\Omega_j$ are the frequencies of individual oscillations. In
contrast to mixing of classical oscillations, only these two
combination frequencies appear in the detector output, and as
in the linear regime, intensity of all spectral peaks is
limited by the quantum mechanics of measurement.

The paper is organized as follows. Section II introduces
a model of mesoscopic solid-state detectors used in this work
and gives an explicit description of one practical detector
which realizes this model and can operate in the purely
quadratic or non-linear regime. Section III describes the
two-qubit system and derives explicit equations for the evolution
of the density matrix of this system in the measurement process.
In Section IV we use these equations to calculate output spectral
density of the non-linear detector measuring the two-qubit system
in several situations. Section V provides concluding remarks.

\section{Mesoscopic quantum measurements}

In this Section, we describe a generic model of quantum
measurements with a mesoscopic solid-state detector and
provide a detailed discussion of one specific example of such a
detector.

\subsection{General model of a mesoscopic detector}

\begin{figure}[htb]
\setlength{\unitlength}{1.0in}
\begin{picture}(3.,1.4)
\put(.1,.1){\epsfxsize=2.8in\epsfbox{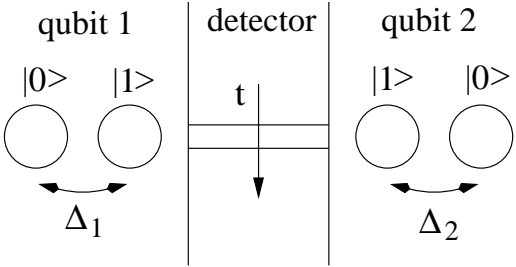}}
\end{picture}
\caption{Diagram of a mesoscopic detector measuring two qubits.
The qubits modulate amplitude $t$ of tunneling of detector
particles between the two reservoirs.} \label{fig1} \end{figure}

Although mesoscopic detectors can have quite different
physical implementations and include, e.g., quantum point
contacts (QPC) \cite{q1,q2,q3,q4,q5,q6,q7}, normal and
superconducting SET transistors \cite{s1,s2,s3,s4,s5,s6,s7},
SQUID magnetometers \cite{b2} and generic mesoscopic conductors
\cite{m1,m2}, the operating principle of all these detectors is
essentially the same. Measured quantum system controls, through
an operator $x$, the transmission amplitude $\hat{t}(x)$ of
particles (they can be electrons, Cooper pairs, or magnetic flux
quanta) between the two reservoirs. The flux of these particles
provides then the information on the state of this system.
Schematics of the detector of this type measuring two qubits is
shown in Fig.~\ref{fig1}.

Quantitatively, the Hamiltonian of the detector-system coupling
for such a detector consists of detector tunneling modulated by
the measured system and can be written as
\begin{equation}
H_T=\hat{t}(x) \xi+\hat{t}^{\dagger} (x) \xi^{\dagger} \, ,
\label{e1} \end{equation}
where $\xi, \xi^{\dagger}$ are the detector operators that create
excitations when a particle is transferred, respectively, forward
and backward between the reservoirs. For the quantum-point-contact
(QPC) detector, which represents the simplest realization of the
model discussed in this Section, $\xi, \xi^{\dagger}$ describe
excitation of electron-hole pairs in the QPC electrodes.

We make several additional assumptions about the detector. We
suppose that the tunneling between the detector reservoirs is weak
and can be accounted for in the lowest non-vanishing order in the
tunneling Hamiltonian (\ref{e1}). Under this assumption, the
precise form of the internal detector Hamiltonian $H_d$ is not
important and dynamics of measurement is defined by the
correlators:
\begin{equation}
\gamma_+=\int_{0}^{\infty} dt\langle \xi(t) \xi^{\dagger}\rangle
\, , \;\;\; \gamma_-=\int_{0}^{\infty}dt\langle \xi^{\dagger}(t)
\xi\rangle \, , \label{e3} \end{equation} where the angled
brackets denote averaging over the detector reservoirs which are
taken to be in a stationary state with the density matrix
$\rho_D$:  $\langle ... \rangle = \mbox{Tr}_D \{ ... \rho_D \}$.
The correlators (\ref{e3}) set the scale $\Gamma_{\pm}\equiv
2\mbox{Re} \gamma_{\pm}$ of the forward and backward detector
tunneling rates. The correlators $\langle \xi(t) \xi \rangle$,
$\langle \xi^{\dagger} (t) \xi^{\dagger} \rangle$ that do not
conserve the number of tunneling particles are assumed to vanish.
As can be seen more explicitly from the example of the SQUID
detector discussed below, this condition holds even in the case of
superconducting detectors, where such ``anomalous'' correlators
can exists in general. Under the condition of large bias voltage
important in our model of the detector operation, the Cooper-pair
tunneling is incoherent, and anomalous correlators indeed vanish.

Another assumption is that the characteristic time of the detector
tunneling is much shorter than that of the evolution of the
measured system. One of the conditions implied by this assumptions
is that the energy bias $\Delta E$ for tunneling through the
detector, which sets one of the tunneling time scales, $\Delta
E^{-1}$, is much larger than the typical energies $E_0$ of the
measured system. In the example of the QPC detector, $\Delta
E=eV$, and this condition means that the bias voltage $V$ across
the QPC is sufficiently large. For short tunneling times, the
functions $\xi(t)$, $\xi^{\dagger}(t)$ in Eq.~(\ref{e3}) are
effectively $\delta$-correlated on the time scale of the dynamics
of the measured system. Condition of the large energy bias $\Delta
E$ for the detector tunneling leading to the correlators
Eq.~(\ref{e3}) being $\delta$-correlated is the natural part of
the measurement model: it enables one to neglect quantum
fluctuations in the detector in the frequency range that
corresponds to that of the measured system, and makes the detector
response in this frequency range classical.

Combined with the assumption of weak tunneling, vanishing
correlation time in the correlators (\ref{e3}) makes it possible
to write down simple evolution equations for the density matrix
$\rho$ of the measured system. Indeed, the time evolution of
$\rho$ in the interaction representation with respect to the
tunneling Hamiltonian (\ref{e1}) is given by the standard
expression:
\begin{equation}
\rho(t) =\mbox{Tr}_D \{ S \rho \rho_D S^{\dagger} \} , \;\; S = T
\exp \{-i\int^tdt' H_T (t') \} . \label{e5} \end{equation} If the
detector operators in Eq.~(\ref{e3}) are $\delta$-correlated, one
can keep only the ``non-crossing diagrams'' in the perturbation
expansion of Eq.~(\ref{e5}) in $H_T$. Evolution of $\rho(t)$ in
Eq.~(\ref{e5}) is governed then by the following equation:
\[ \dot{\rho}= \Gamma_+\hat{t}^{\dagger}\rho \hat{t}+\Gamma_-
\hat{t} \rho \hat{t}^{\dagger} -(\gamma_+\hat{t}\hat{t}^{\dagger}
+ \gamma_-\hat{t}^{\dagger}\hat{t})\rho - \]

\vspace{-2ex}

\begin{equation}
\rho(\gamma^*_+\hat{t} \hat{t}^{\dagger}+ \gamma^*_-
\hat{t}^{\dagger}\hat{t}) \, .
\label{e6} \end{equation}
Equation (\ref{e6}) describes the measurement-induced part of
the evolution of an arbitrary measured system within our generic
model of a mesoscopic detector.

One general remark that should be made here is that for some
detectors, e.g. SET transistors, there are regimes of operation,
when the particle transfer through the detector consists of more
than one steps and can not be characterized by one transmission
amplitude. Our measurement model is not applicable in these
regimes. We focus on the case of one-step transfer, however, since
only in this case the detector can be quantum-limited:
intermediate steps introduce additional back-action dephasing for
the measured system without increasing information contained in
the detector tunneling rate.

Evolution (\ref{e6}) of the density matrix $\rho$ of the measured
system is reflected in the detector output: the particle current
between the detector reservoirs. Using the same logic that lead to
Eq.~(\ref{e6}), one can see that the assumed large difference
between the time scale of detector tunneling and evolution of the
measured system makes it possible to reduce the expression for the
particle current to the form of the operator $I$ in the space of
the measured system:
\begin{equation}
I= (\Gamma_+ -\Gamma_-)t^{\dagger}t\, .
\label{ea1} \end{equation}
This equation gives both the dc current
\begin{equation}
\langle I\rangle =\mbox{Tr}\{I \rho_0 \} \, ,
\label{ea2} \end{equation}
where $\rho_0$ is the stationary solution of Eq.\ (\ref{e6}),
and the current spectral density
\begin{equation}
S_I = S_0 + 2\int_0^{\infty} d \tau \cos\omega \tau ( \mbox{Tr}\{
Ie^{L\tau}[ I\rho_0] \} -\langle I\rangle^2)\, . \label{ea3}
\end{equation} Here $S_0$ represents the usual noise (typically, a
mixture of shot and thermal noise) associated with tunneling:
\begin{equation}
S_0 = (\Gamma_+ +\Gamma_-)\mbox{Tr}\{ t^{\dagger}t\rho_0 \}\, ,
\label{ea4} \end{equation}
and $e^{L\tau} [A]$ denotes the evolution of the matrix $A$
during time interval $\tau$ governed by Eq.\ (\ref{e6}).

\subsection{DC SQUID as the non-linear detector}

\begin{figure}[htb]
\setlength{\unitlength}{1.0in}
\begin{picture}(3.,1.85)
\put(.55,.1){\epsfxsize=2.2in\epsfbox{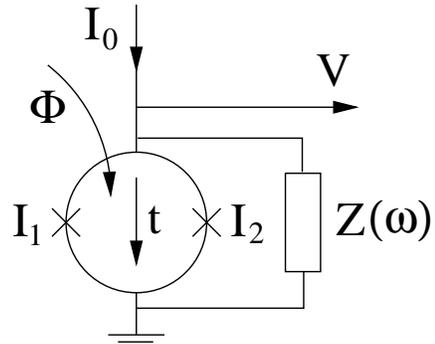}}
\end{picture}
\caption{DC SQUID detector formed by two Josephson junctions with
critical currents $I_{1,2}$ in
a small superconducting loop shunted by the impedance $Z(\omega)$.
The SQUID is biased by the dc current $I_0$, the voltage $V$ is
the measurement output and reflects the variations in the rate of
incoherent Cooper-pair transfer through the SQUID with amplitude
$t$ controlled by the flux $\Phi$. Existence of the two
interfering contributions to the tunneling amplitudes makes it
possible for the SQUID to operate as a quadratic detector. }
\label{fig2} \end{figure}

Before discussing applications of our measurement model that
is expressed quantitatively by Eqs. (\ref{e6})--(\ref{ea4}),
we give one explicit example of the mesoscopic detector
(different from the simplest QPC detector that was already
mentioned several times in this Section) that realizes this
model. The detector is the strongly-biased dc SQUID
shown schematically in Fig.~\ref{fig2}.
It consists of two Josephson junctions with critical currents
$I_{1,2}$ that are included in small superconducting loop
shunted by a resistor with some impedance $Z(\omega)$,
assumed to be featureless, $Z(\omega)=Z(0)\equiv R$, at low
frequencies of variations of the measured magnetic flux $\Phi$
through the SQUID loop. When the loop inductance is small, the
difference between the two Josephson phases $\varphi_{1,2}$
across the two junctions is directly linked to $\Phi$:
\[ \varphi_{1}-\varphi_{2}=2\pi \Phi/\Phi_0 \equiv \phi \, , \]
where $\Phi_0=\pi\hbar/e$ is the magnetic flux quantum. In this
regime, the SQUID is equivalent to a single Josephson junction,
with the effective amplitude of Cooper-pair tunneling controlled
by the flux $\Phi$. Due to interference of the Cooper-pair
tunneling in the two SQUID junctions, the total tunneling
amplitude is equal to the sum of tunneling amplitudes in the two
Josephson junctions, with the amplitude in each junction being
proportional to its critical current $I_j$. The total amplitude
can then be written as
\begin{equation}
t(\phi) = i_1e^{i\phi/2} + i_2e^{-i\phi/2} \, ,
\label{e7} \end{equation}
where $i_1=(I_1/I_2)^{1/2}$ and $i_2=1/i_1$ characterize the
asymmetry of the two junctions.

The operating principle of the SQUID as a detector coincides with
the one discussed in the beginning of this Section: variations of
the flux $\phi$ around some bias point $\bar{\phi}$ lead to
variations of the Cooper-pair tunneling amplitude (\ref{e7})
affecting the transfer rate $I$ of Cooper pairs through the SQUID.
Cooper-pair tunneling rate is reflected in the detector output:
deviations $V$ of the voltage across the SQUID, $V=-2eIR$, from
the value $V_0=RI_0$ induced by the dc current bias $I_0$ (see
Fig.~\ref{fig2}). An important feature of the SQUID detector is
that in the case of identical junctions: $i_1=i_2=1$, at the bias
points when $\bar{\phi}/2\pi$ is integer, the tunneling amplitude
$t(\phi)$ (\ref{e7}) varies quadratically as a function of $\phi$,
with vanishing coefficient of the linear response. As shown
explicitly in the next Section, under this bias condition the
SQUID can act as the purely quadratic detector.

The Cooper-pair tunneling Hamiltonian in the SQUID can be
expressed through the tunneling amplitude (\ref{e7}) similarly
to Eq.~(\ref{e1}):
\begin{equation}
H_T= -\frac{(I_1I_2)^{1/2}}{4e} [t(\phi)e^{i(2eV_0t+\varphi(t)
)}+\hat{t}^{\dagger} (\phi) e^{-i(2eV_0t+\varphi(t))} ] \, ,
\label{e8} \end{equation}
where $\varphi(t)=(\varphi_{1}(t)+\varphi_{2}(t))/2$ is the
average Josephson phase across the SQUID which includes
fluctuating component accumulated due to the equilibrium
voltage fluctuations across the impedance $Z(\omega)$. The
fluctuations couple the Cooper-pair tunneling to $Z(\omega)$ so
that the transfer of a Cooper pair across the SQUID creates
electromagnetic excitations in this impedance, the process
described by operators $e^{\pm i \varphi(t)}$ in Eq.~(\ref{e8}).
In this way, impedance $Z(\omega)$ provides dissipation
necessary to make the Cooper-pair transfer irreversible.

When the dc bias current $I_0$ and associated voltage
$V_0=RI_0$ across the dc SQUID are sufficiently large, one can
treat the tunnel Hamiltonian (\ref{e8}) as perturbation (i.e., the
dc current through the SQUID is much smaller than the bias current
$I_0$) and the system dynamics at low frequencies can be described
as incoherent tunneling of Cooper pairs through the SQUID
\cite{b10}. This regime satisfies both of the conditions on the
detector dynamics discussed in the general detector model, and
when the control flux $\phi$ is created by the measured quantum
system, measurement dynamics is governed by the same general
equation for the system density matrix (\ref{e6}), where the
correlators (\ref{e3}) are now given by the following expressions
\cite{b10}:
\begin{equation}
\gamma_{\pm} = \frac{I_1I_2}{16e^2} \int_{0}^{\infty} dt e^{ \pm
i2eV_0t} \exp \{ \int \frac{d \omega }{\omega} \frac{\mbox{Re}
Z(\omega) }{ R_Q}\frac{ e^{- i\omega t } -1 }{1-e^{- \omega /T} }
\} \, . \label{e9} \end{equation} Here $R_Q \equiv \pi \hbar/4e^2$
is the ``quantum resistance''. Under the conditions of weak linear
coupling of the SQUID detector to the measured systems, the
correlators (\ref{e9}) that give the rates of incoherent
Cooper-pair tunneling are the only relevant properties of the
SQUID operating as a linear quantum detector \cite{b11}. If the
variations of the external flux $\phi$ controlling Cooper-pair
tunneling are not small, the SQUID can act as non-linear, and in
particular, purely quadratic detector. Dynamics of the measured
system is reflected in the Cooper-pair current in the SQUID in
exactly the same way as in the general model [see
Eqs.~(\ref{ea1})--(\ref{ea4})], and is converted into the detector
output voltage $V$ as discussed above.

\section{Two qubits measured continuously by the non-linear
detector}

The main focus of this work is on the measurement of the system
of two, in general interacting, qubits. The qubit Hamiltonian is:
\begin{equation}
H_0 = -\frac{1}{2} \sum_{j=1,2} (\varepsilon_j \sigma_{z}^{j} +
\Delta_j \sigma_x^{j}) + \frac{\nu}{2} \sigma_{z}^{1}
\sigma_{z}^{2} \, , \label{e2} \end{equation} where $\Delta_j$ is
the tunnel amplitude and $\varepsilon_j$ is the bias of the $j$-th
qubit ($j=1,2$), $\nu$ is the qubit-qubit interaction energy, and
$\sigma$'s here and below denote the Pauli matrices. The qubits
are coupled to one detector (see Fig.~\ref{fig1}) through their
basis-forming variables $\sigma_z^j$, and the control operator $x$
modulating the detector tunneling (\ref{e1}) is
$x=c_1\sigma_z^1+c_2 \sigma_z^2$. It gives the following
expression for the tunneling amplitude $\hat{t}$:
\begin{equation}
\hat{t}(x) = t_0+\sum_j \delta_j \sigma_z^{j}+ \lambda
\sigma_z^{1} \sigma_z^{2} \, .
\label{e4} \end{equation}
The last term in this expression appears due to non-linearity of
the dependence of the transmission amplitude $t$ on variable $x$.
If the linear terms in Eq.~(\ref{e4}) vanish, $\delta_j=0$, while
$\lambda \neq 0$, one has purely quadratic detector. In the case
of two qubits, since the Pauli matrices satisfy the condition
$\sigma^2=1$, Eq.~(\ref{e4}) represents the most general
dependence of $t$ on $\sigma_z^{j}$.\cite{rem2} In the example of
the SQUID detector discussed in Sec.~II, when the two flux qubits
are coupled to it, the normalized flux $\phi$ through the detector
is:
\[ \phi=\bar{\phi}+2 \sum_j \delta \phi_j\sigma_z^{j} \, , \]
where $\delta \phi_j$ characterizes the strength of coupling to
the $j$th qubit, and the average flux $\bar{\phi}$ sets the
detector operating point. With such a coupling, the amplitude
(\ref{e7}) of the Cooper-pair tunneling in the detector is given
by Eq.~({\ref{e4}) with
\begin{eqnarray}
t_0 = [i_1e^{i\bar{\phi}/2} + i_2e^{-i\bar{\phi}/2} ]
\cos \delta \phi_1 \cos \delta \phi_2, , \nonumber \\
\delta_j=i [i_1e^{i\bar{\phi}/2} -i_2e^{-i\bar{\phi}/2} ]
\cos \delta \phi_{j'} \sin \delta \phi_j\, ,\;\;\; j'\neq j \, ,
\nonumber \\
\lambda  = - [i_1e^{i\bar{\phi}/2} + i_2e^{-i\bar{\phi}/2} ]
\sin \delta \phi_1 \sin \delta \phi_2\, . \nonumber
\end{eqnarray}
If $\bar{\phi}=2\pi n$ with integer $n$, and the SQUID is
symmetric: $i_1=i_2=1$,the linear coupling coefficients
$\delta_j$ vanish and the SQUID has only quadratic response.

For qubit-detector coupling of the form (\ref{e4}), it is
convenient to write the general equation (\ref{e6}) for the
density matrix $\rho$ of the two-qubit system in the
``measurement'' basis of eigenstates of the $\sigma_z^{j}$
operators,  $|\! \uparrow \uparrow \rangle \, , |\! \uparrow
\downarrow \rangle \, ,|\! \downarrow \uparrow \rangle \, ,$
and $|\! \downarrow \downarrow \rangle$. Each state $|k\rangle$
of this basis is characterized by the magnitude $t_k$ of the
transmission amplitude (\ref{e4}):
\begin{eqnarray}
t_1=t_0+ \delta_1 +\delta_2 + \lambda\, ,\;\;\;
t_2=t_0+ \delta_1-\delta_2- \lambda \, , \nonumber \\
t_3=t_0-\delta_1 +\delta_2- \lambda\, ,\;\;\;
t_4=t_0- \delta_1 -\delta_2 + \lambda \, , \nonumber
\end{eqnarray}
and associated value of the detector tunneling current, for
which we use the obvious notations:
\begin{eqnarray*}
I_{\uparrow \uparrow} = (\Gamma_+-\Gamma_-)|t_0+ \delta_1
+\delta_2 + \lambda|^2 \, ,  \\
I_{\uparrow \downarrow}=(\Gamma_+-\Gamma_-)|t_0+ \delta_1
-\delta_2- \lambda|^2 \, ,  \\
I_{\downarrow \uparrow} = (\Gamma_+-\Gamma_-)|t_0- \delta_1
+\delta_2 - \lambda|^2 \, ,   \\
I_{\downarrow \downarrow}=(\Gamma_+-\Gamma_-)|t_0- \delta_1
-\delta_2 + \lambda|^2 \, .
\end{eqnarray*}
Combining measurement-induced evolution (\ref{e6}) with
the evolution due to the qubit Hamiltonian $H_0$ we get the
equation for $\rho$ in the measurement basis:
\begin{eqnarray}
\dot{\rho}_{kl}= -\gamma_{kl}\rho_{kl} -i[H_0,\rho]_{kl}\, ,
\label{e10} \\ \gamma_{kl} =(1/2)(\Gamma_+ +\Gamma_-)
|t_k-t_l|^2 \, . \nonumber
\end{eqnarray}
In Eq.~(\ref{e10}), the Hamiltonian $H_0$ includes two
renormalization terms:
\[ H_0 \rightarrow H_0 +\delta H+\delta H'\, . \]
The first one is due to imaginary parts of the correlators
(\ref{e3}):
\begin{equation}
\delta H=\sum_j \delta \varepsilon_j\sigma_z^{j}+ \delta \nu
\sigma_z^{1}\sigma_z^{2}
\label{e11} \end{equation}
where
\begin{eqnarray}
\delta \varepsilon_j &=& \mbox{Re} (\delta_jt_0^* + \delta_{j'}
\lambda^*) \mbox{Im} (\gamma_- +\gamma_+) \, , \nonumber \\
\delta \nu &=& \mbox{Re} (\delta_1\delta_2^* + t_0\lambda^*)
\mbox{Im} (\gamma_- +\gamma_+) \, ,\nonumber
\end{eqnarray}
and $j,j'=1,2$, with $j'\neq j$.
The second term $\delta H'$ is due to phases $\varphi_{kl} \equiv
\arg(t_kt_l^*)$ of the transfer amplitudes $t_k$ and is defined
by the following relation:
\begin{equation}
[\delta H', \rho]_{kl}= (\Gamma_+ -\Gamma_-)|t_kt_l|
\sin \varphi_{kl}\rho_{kl}\, .
\label{e12} \end{equation}
Note that $\delta H'$ can not always be cast in the form
(\ref{e11}) in which it can be absorbed in the renormalization
of qubit energies in the Hamiltonian (\ref{e2}). This can be
done if the difference between the tunneling amplitudes $t_k$
are small, $|t_k-t_l| \ll |t_k|$. In this work, we will
assume that this condition on $\delta H'$ is satisfied, and
the renormalized Hamiltonian has the same form (\ref{e2}).

Equation (\ref{e10}) describes the time evolution of the
qubits averaged over different measurement outcomes,
which in our measurement model are represented by the number
$n$ of particles tunneled through the detector. Because of
this averaging, the qualitative effect of measurement in
Eq.~(\ref{e10}) is the ``back-action'' dephasing of different
states of the measurement basis with the rates $\gamma_{kl}$.
In general, Eq. (\ref{e6}) contains information about the
dynamics of $n$ and can be used to write down the evolution
equations conditioned on the specific measurement outcomes
(see Ref.~\onlinecite{b12}). The averaged equation (\ref{e10})
is sufficient for calculation of the output spectrum of the
detector which is the purpose of this work.

The assumption of short detector tunneling time that lead
to Eq.~(\ref{e10}) makes this equation valid even in the regime
when the detector-qubit coupling is strong and the dephasing
rates $\gamma_{kl}$ are large in comparison with the rates of
evolution due to the Hamiltonian $H_0$. Straightforward numerical
solution of Eq.~(\ref{e10}) combined with Eqs.~(\ref{ea3}) and
(\ref{ea4}) gives the spectrum of the detector output. (All
numerical plots of the spectra are obtained below in this way.)
An example of such a spectrum for most generic set of parameters
of the qubit-detector system is shown in Fig.~\ref{fig3}. For
these parameters all six intervals between four energy levels
of the two-qubit Hamiltonian (\ref{e2}) are different, and show
up as six finite-frequency peaks in the spectrum of the detector
output. There is also the zero-frequency peak that reflects the
dynamics of detector-induced transitions between the energy
levels. Qualitatively, the width of all peaks corresponds to the
rates of transitions between the energy levels, with finite
frequency peaks are also broadened by the ``pure dephasing''
terms in the evolution equation for the density matrix of the
system [see Eq.~(\ref{e13}) below] that are not related to the
transitions.

Although so far the nonlinear mesoscopic detectors of the type 
considered in this work have not been realized experimentally, 
many elements of our detector model were indeed demonstrated. For 
instance, the QPC detector, which is the simplest realization of 
this model, was used to measure coherent oscillations in a quantum-dot 
qubit \cite{q8}. Parameters of the detector-qubit system used in 
Fig.3 and other numerical results presented below (in particular the 
characteristic detector tunneling rate $\Gamma_+|t_0|^2$ on the order 
of the qubit oscillation frequency $\Delta$) are consistent with 
those in experiments. 

\begin{figure}[htb]
\setlength{\unitlength}{1.0in}
\begin{picture}(3.,2.1)
\put(.1,.0){\epsfxsize=2.7in\epsfbox{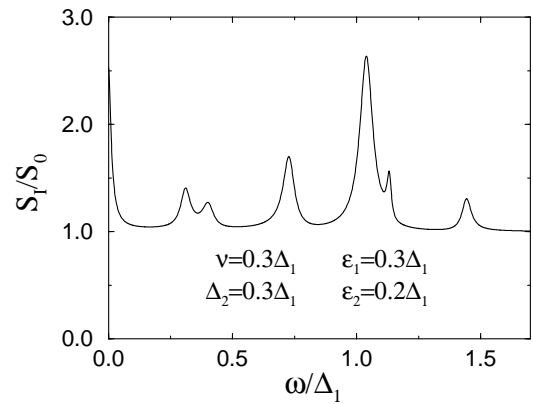}}
\end{picture}
\caption{Output spectrum of a nonlinear detector measuring
two qubits with ``the most general'' set of parameters. Six peaks
in the spectrum at finite frequencies correspond to six different
energy intervals in the energy spectrum of the two-qubit system.
The zero-frequency peak reflects dynamics of transitions between
energy levels. Detector parameters are: $\delta_1=0.1,\,
\delta_2=0.07,\, \lambda=0.09$ (all normalized to $t_0$).
In this Figure, and in all numerical plots below we take
$\Gamma_+|t_0|^2=\Delta_1$, $\Gamma_-=0$, and assume that
the detector tunneling amplitudes are real.}
\label{fig3} \end{figure}

In the situation without any symmetries in the qubit parameters,
matrix elements of the detector-qubit coupling are non-vanishing
between all four energy eigenstates of the two-qubit system
(\ref{e2}). Moreover, the assumption of the short detector
tunneling time implies that the transition rate is independent of
the energies of these states: i.e., if one views the detector as
the reservoir producing dephasing for the measured system, the
effective reservoir temperature is much larger than the system
energies. This means that the stationary density matrix of the
system is:
\begin{equation}
\rho_0 =\hat{I}/4 \, ,
\label{e14} \end{equation}
where $\hat{I}$ is the unity matrix in the four-dimensional
space of the two-qubit system. This means that the dc detector
current (\ref{ea2}) in this regime is:
\begin{eqnarray}
\langle I\rangle =(I_{\uparrow \uparrow}+I_{\uparrow
\downarrow}+ I_{\downarrow \uparrow} + I_{\downarrow
\downarrow})/4 \nonumber \\ = (\Gamma_+ -\Gamma_-)
(|t_0|^2+|\delta_1|^2+|\delta_2|^2+ |\lambda|^2)\, ,
\label{e15} \end{eqnarray}
and associated noise $S_0$ (\ref{ea4}) is given by the same
expression with $\Gamma_+ -\Gamma_-$ replaced by $\Gamma_+
+\Gamma_-$. If, however, the qubit parameters have some
symmetry (examples are given in the Sections that follow),
matrix elements satisfy ``selection rules'' and vanish for
transitions into some states. The stationary density matrix
$\rho_0$ is then proportional to the unity matrix confined to
a smaller subspace of the Hilbert space of the two-qubit system.
In this case the dc current $\langle I\rangle$ and the noise
$S_0$ are given by equations different from Eq.~(\ref{e15}),
and the number of peaks in the detector spectrum is less than
the maximum of 6 peaks.

It is interesting to note that the selection rules that
determine the number of finite-frequency peaks in the output
spectrum of the detector allow sometimes for a simple qualitative
classical interpretation. For instance, as will be demonstrated
more rigorously below, in agreement with the case of classical
oscillations, if the qubit-qubit interaction vanishes, and the
detector is purely linear, the output spectrum contains at most two
peaks that correspond to oscillations in the individual qubits with
frequencies $\Omega_{1,2}$. If the detector is non-linear, the
spectrum acquires two more peaks at the combination frequencies
$\Omega_1 \pm \Omega_2$. The maximum total number of peaks is also
equal to four if the detector is linear but the qubit-qubit
interaction is finite.

Quantitatively, the detector spectrum can be obtained from the
solution of Eq.~(\ref{e10}) for evolution of the qubit density
matrix. Despite the relative simplicity of its numerical solution,
Eq.~(\ref{e10}) can be solved analytically only in a few cases [see,
e.g., Eq.~(\ref{e20}) below]. More detailed analytical results
can be obtained if the dephasing rates $\gamma_{kl}$ are small in
comparison to the intervals $\omega_{nm}\equiv E_n-E_m$ between
eigenenergies of the Hamiltonian $H_0$ (\ref{e2}). In this
limit, it is convenient to transform Eq.~(\ref{e10}) into the
basis of eigenstates $|n\rangle$ of $H_0$, $H_0|n\rangle= E_n
|n\rangle$, where one can separate components of the density
matrix $\rho$ that evolve with different frequencies. Then,
making use of the standard approach (see, e.g., \cite{b13})
equation for $\rho$ can be simplified by neglecting the terms
that mix these components. Written in the energy eigenstate basis,
Eq. (\ref{e10}) is reduced in this way to the following form:
\begin{eqnarray}
\dot{\rho}_{nm} = -i\omega_{nm}\rho_{nm} +(\Gamma_+ +\Gamma_-)
\times \nonumber \\ \Big[ - \Big(\sum_{p\neq m} |\hat{t}_{m p}
|^2+\sum_{p\neq n} |\hat{t}_{np}|^2  +  |\hat{t}_{mm}-
\hat{t}_{nn}|^2 \Big) \rho_{nm}/2 + \nonumber \\ \delta_{nm}
\sum_{p} \rho_{pp} |\hat{t}_{mp}|^2 + (1-\delta_{nm})
\sum_{(p,q)} \rho_{pq} \mbox{Re} (\hat{t}^{\dagger}_{np}
\hat{t}_{qm} ) \Big]\, .
\label{e13} \end{eqnarray}
where $\hat{t}_{pq}$ are the matrix elements of the tunneling
amplitude $\hat{t}$ (as operator in the qubit space) in the basis
of energy eigenstates, and the last sum is taken over the pairs
$(p,q)$ of states that satisfy the ``resonance'' condition:
$E_p-E_q= E_n-E_m \, , \; (p,q) \neq (n,m)$, and therefore
contribute to the same spectral peak of the detector output.
Solution of Eq.~(\ref{e13}) can be used to calculate the average
detector current and its spectrum using the same
Eqs.~(\ref{ea1})--(\ref{ea4}). In the next Section, we use
both Eq.~(\ref{e10}) and Eq.~(\ref{e13}) to describe the time
evolution of the two-qubit system and resulting detector output
in different situations.

\section{Spectral density of the detector output} 

Dynamics of the detector-qubit system and spectral density of 
the detector output depend on several characteristics of the 
system, most important of which are: the degree of non-linearity 
of the detector-qubit coupling, symmetry of parameters of 
individual qubits, and strength of the qubit-qubit interaction. 
In this Section, we calculate the spectra of the detector output 
in several regimes that differ in terms of these characteristics. 
Subsection A discusses the case of purely quadratic coupling of 
the detector to qubits that, in general, are different and 
interacting. Subsection B deals with the detector-qubit coupling 
which includes both the linear and quadratic terms but only for 
identical qubits. In the last Subsection C, we consider the most 
general case of different and interacting qubits measured by the 
detector with arbitrary non-linearity limiting ourselves to the 
regime of weak detector-qubit coupling.

\subsection{Purely quadratic detector}
We begin the discussion of the detector output spectrum by
considering the purely quadratic measurement of {\em unbiased}
qubits ($\delta_j=\varepsilon_j=0$). In this case, dynamics of
measurement governed by Eq.~(\ref{e10}) is such that the Hilbert
space of the two-qubit system is split into two two-dimensional
subspaces $D_{\pm}$ with no transitions between them, so that the
system evolves independently in each subspace. If the basis of
states in the subspace $D_{\pm}$ is chosen as
\begin{equation}
D_{\pm} \equiv  \{ \frac{1}{\sqrt{2}} (|\!\uparrow \uparrow
\rangle \pm |\!\downarrow \downarrow \rangle) \, ,
\frac{1}{\sqrt{2}} (|\!\downarrow \uparrow \rangle \pm |\!\uparrow
\downarrow \rangle) \} \, , \label{e16}
\end{equation} the Hamiltonian (\ref{e2}) is split into two
independent parts $H_{\pm}$ acting within $D_{\pm}$:
\begin{equation}
H_{\pm} = -(1/2) [(\Delta_1\pm \Delta_2) \bar{\sigma}_x +
\nu \bar{\sigma}_z ] \, .
\label{e17} \end{equation}
(By putting the bars on the Pauli matices $\sigma$ in this
equation, we want to distinguish these Pauli matrices acting
in the subspaces $D_{\pm}$ from those describing individual
qubits.)

Qualitatively, the qubit dynamics within each subspace can be
viewed as oscillations between parallel and anti-parallel
configurations. The oscillation frequencies are equal to
$\Delta_1\pm \Delta_2$ for non-interacting qubits and are
shifted upwards by interaction which creates energy difference
between the parallel and anti-parallel configurations.
Since the purely quadratic detector distinguishes these two
types of qubit configurations, the current operator (\ref{ea1})
within each subspace in the basis (\ref{e16}) is:
\begin{equation}
I= \langle I \rangle + I_a \bar{\sigma}_z \, ,
\label{e18} \end{equation}
where the average detector current $\langle I \rangle$ is given
by Eq.~(\ref{e15}) with $\delta_j =0$, and
\[ I_a = (\Gamma_+ -\Gamma_-)2\mbox{Re} (t_0\lambda^*) =
I_{\uparrow \uparrow}-I_{\uparrow \downarrow} \]
has the meaning of the amplitude of the detector current
oscillations reflecting the qubit oscillations between the
parallel and anti-parallel configurations.

Quantitatively, the evolution equation (\ref{e10}) within
$D_{\pm}$ is:
\begin{equation}
\dot{\rho}^{(\pm)}_{kl} = -i[H_{\pm},\rho^{(\pm)}]_{kl} -
\Gamma \left( \begin{array}{cc} 0\, , & \rho_{12}^{(\pm)}  \\
\rho_{21}^{(\pm)}\, , & 0 \end{array} \right) \, ,
\label{e19} \end{equation}
where $\Gamma \equiv 2 (\Gamma_+ +\Gamma_-)|\lambda|^2$ is
the measurement-induced dephasing of the basis states
(\ref{e16}) of each subspace. For {\em non-interacting}
qubits, solving this equation and calculating the spectral
density of the detector output according to Eq.~(\ref{ea3}) we
get \cite{b12}:
\begin{equation}
S_I^{\pm} (\omega ) = S_0 + \frac{2 I_a^2 (\Delta_1\pm
\Delta_2)^2 \Gamma} {(\omega^2-(\Delta_1\pm \Delta_2)^2)^2+
\Gamma^2\omega^2} \, ,
\label{e20} \end{equation}
where $S_0=(\Gamma_+ +\Gamma_-)(|t_0|^2+|\lambda|^2)$.

The two spectral densities (\ref{e20}) correspond to two possible
outcomes of measurement: the qubits found in one or the other
subspace $D_{\pm}$, the probability of the outcomes being
determined by the initial state of the qubits. Each of the
spectral densities coincides with the spectral density of the
linear detector measuring coherent oscillations in one qubit
\cite{b3}. Similarly to that case, the maximum of the ratio of the
oscillation peak versus noise $S_0$ for each spectrum $S_I^{\pm}
(\omega )$ is 4. As one can see from Eq.\ (\ref{e20}), this
maximum is reached when the measurement is weak: $|\lambda|\ll
|t_0|$, and the detector is ``ideal'': $\arg(t_0\lambda^*)$=0, and
only $\Gamma_+$ or $\Gamma_-$ is non-vanishing. If, however, there
is small but finite transition rate between the two subspaces that
mixes the two outcomes of measurement, the peak height is reduced
by averaging over the two spectral densities (\ref{e20}). This
situation is illustrated in Fig.~\ref{fig4} which shows the output
spectra of the purely quadratic detector, when the subspaces
$D_{\pm}$ are mixed by small qubit bias $\varepsilon$. Since the
stationary density matrix (\ref{e14}) is equally distributed over
all qubit states, the two peaks of the spectral densities
(\ref{e20}) are mixed with equal probabilities, and the maximum of
the ratio of the oscillation peak heights versus noise $S_0$ for
the combined spectrum $S_I (\omega )$ is 2. Spectrum shown in
Fig.~\ref{fig4} for $\varepsilon=0.1\Delta_1$ (solid line) is
close to this limit.

\begin{figure}[htb]
\setlength{\unitlength}{1.0in}
\begin{picture}(3.,2.)
\put(.1,.0){\epsfxsize=2.7in\epsfbox{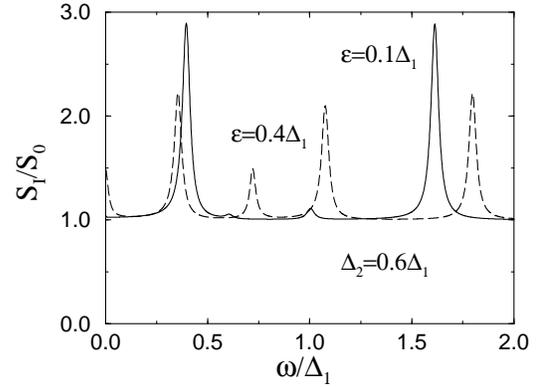}}
\end{picture}
\caption{Output spectra of a purely quadratic detector
measuring two non-interacting qubits. Small qubit bias
$\varepsilon_1 =\varepsilon_2 \equiv \varepsilon$ (solid
line) creates transitions that lead to averaging of the
two main peaks at combination frequencies $\Delta_1\pm
\Delta_2$ [see \protect Eq.~(\ref{e20})]. Further increase
of $\varepsilon$ (dashed line) makes additional spectral
peaks associated with these transitions more pronounced.
The strength of quadratic qubit-detector coupling is taken
to be $\lambda=0.15t_0$.}
\label{fig4} \end{figure}

For {\em interacting} qubits, analytical results for the
spectrum can be obtained in the case of weak back-action
dephasing $\Gamma \ll |\Delta_1\pm \Delta_2|$, when the
evolution equations (\ref{e19}) within each subspace can
be transformed into the energy eigenstates representation
and simplified there similarly to Eq.~(\ref{e13}). In this
limit, the shape of all spectral peaks is Lorentzian, with
the decay rate $\gamma$ of the off-diagonal elements of the
density matrices $\rho^{(\pm)}$ in the eigenstate
representation,
\[ \gamma = (\Gamma_+ +\Gamma_-)|\lambda|^2 (1+\nu^2/
\Omega_{\pm}^2) \, , \]
determines the width of the peaks at the combination
frequencies shifted by interaction:
\begin{equation}
\Omega_{\pm} = [(\Delta_1\pm \Delta_2)^2+\nu^2]^{1/2} \, .
\label{e49} \end{equation} In the case of non-vanishing
interaction strength $\nu$, the eigenstates of the Hamiltonians
(\ref{e17}) have different amplitudes of states (\ref{e16}) with
parallel and anti-parallel qubit configurations and therefore have
different values of the average detector current (\ref{e18}). This
means that for $\Gamma \ll |\Delta_1\pm \Delta_2|$, nonvanishing
interaction $\nu$ creates the spectral peak at zero frequency
which reflects the transitions between the energy eigenstates. The
width of this peak $\gamma_0$ is determined by the rate of
transitions between these states:
\[ \gamma_0 = 2(\Gamma_+ +\Gamma_-)|\lambda|^2
(\Delta_1\pm \Delta_2)^2 / \Omega_{\pm}^2 \, . \]
The total spectra $S_I^{\pm} (\omega)$ within each subspace
$D_{\pm}$ are:
\begin{eqnarray}
S_I^{\pm}(\omega )=S_0 + \frac{I_a^2 }{\Omega_{\pm}^2}
\left( \frac{2 \nu^2 \gamma_0}{\omega^2+ \gamma_0^2}
\right. \nonumber \\  +\left.
\big[ \frac{(\Delta_1 \pm \Delta_2)^2\gamma } {(\omega -
\Omega_{\pm} )^2+\gamma^2 } +
(\omega \rightarrow -\omega) \big] \right) \, .
\label{e27} \end{eqnarray}
Finite qubit bias should lead to averaging of the two spectra
$S_I^{\pm}$ (\ref{e27}) similar to that discussed in the case
of non-interacting qubits and illustrated in Fig.~\ref{fig4}.

If the difference of the qubit tunneling amplitudes become
small, $|\Delta_1- \Delta_2|\ll \Gamma$, qubit dynamics in the
$D_-$ subspace can no longer be viewed as coherent oscillations
between the basis states (\ref{e16}), but rather as incoherent
transitions between these states. The rate of these incoherent
transition $\tau^{-1}$ can be found by treating $\Delta_1-
\Delta_2$ as perturbation in the evolution equation (\ref{e19})
for the density matrix $\rho^{(-)}$:
\begin{equation}
\tau^{-1}= \frac{1}{2} \frac{(\Delta_1- \Delta_2)^2 \Gamma
}{\nu^2 + \Gamma^2} \, .
\label{e28} \end{equation}
Since the basis states (\ref{e16}) are characterized by
different values $\langle I \rangle + I_a$ of the detector
current -- see Eq.~(\ref{e18}), these transition give rise
to the spectral peak in the detector output spectrum at
zero-frequency:
\begin{equation}
S_I^{-} (\omega )=S_0 + \frac{2I_a^2 \tau}{1+\tau^2\omega^2 }
\, , \;\;\; \omega \simeq \tau^{-1} \, .
\label{e29} \end{equation}
Equation (\ref{e29}) for the zero-frequency peak in the
detector output is valid for arbitrary relation between the
back-action dephasing rate and interaction strength $\nu$.
It reproduces both the zero-frequency peak in the spectrum
(\ref{e27}) when $\Gamma \ll \nu$, and if $\Gamma \gg \nu$,
the peak in $S_I^{-}$ obtained from Eq.~(\ref{e20}) in the
limit $|\Delta_1- \Delta_2|\ll \Gamma$. If the qubits are
identical, $\Delta_1=\Delta_2$, the transitions (\ref{e28})
are suppressed and both basis states of the subspace $D_-$
represent separate outcomes of quadratic measurement.

\subsection{Non-linear detector measuring identical qubits}

In the case when the two qubits have the same set of parameters,
it is convenient to discuss dynamics of the two-qubit system
using the language of the total ``spin'' $S=(\sigma_1+\sigma_2)/2$.
The qubit Hamiltonian (\ref{e2}) and the detector tunnel amplitude
(\ref{e4}) providing the qubit-detector coupling can be written
in terms of $S$ like this:
\begin{eqnarray}
H_0 = -\varepsilon S_z- \Delta S_x + \nu S_z^2 \, ,
\label{e30} \\
t =t_0+2 \delta S_z + \lambda (2 S_z^2-1)\, ,
\label{e31} \end{eqnarray}
where $\varepsilon$, $\Delta$, $\delta$ without indices denote
the same quantities as for individual qubits. The state
$(|\!\downarrow \uparrow \rangle - |\!\uparrow \downarrow
\rangle)/\sqrt{2}$ with $S=0$ does not evolve in
time under the Hamiltonian (\ref{e30}) and represents one of
the measurement outcomes characterized by the dc detector
current
\[ \langle I\rangle=(\Gamma_+ -\Gamma_-)|t_0-\lambda|^2 \]
and flat output spectrum $S_I(\omega)=(\Gamma_+ +\Gamma_-)
|t_0-\lambda|^2$.

Three other, $S=1$, states are mixed by measurement and
represent the second measurement outcome. Inserting
Eq.~(\ref{e31}) into Eq.~(\ref{ea1}) and using the fact that
in the $S=1$ subspace $S_z^3=S_z$, we get the following
expression for the detector current operator in this subspace:
\begin{equation}
I =(\Gamma_+ -\Gamma_-) |t_0-\lambda|^2 + a_1 S_z + 2 a_2
S_z^2 \, ,
\label{e34} \end{equation}
where $a_{1,2}$ have the meaning of the amplitudes of current
oscillations between different qubit states:
\begin{eqnarray}
&& a_1=4(\Gamma_+ -\Gamma_-) \mbox{Re} [(t_0+\lambda)\delta^*] =
(I_{\uparrow \uparrow}-I_{\downarrow \downarrow})/2\, ,
\label{e48} \\
&& a_2=2(\Gamma_+ -\Gamma_-) (\mbox{Re} [t_0\lambda^*]+|\delta|^2) =
(I_{\uparrow \uparrow}+I_{\downarrow \downarrow}-2
I_{\uparrow \downarrow})/4\, . \nonumber
\end{eqnarray}
Similarly to Eq.~(\ref{e14}), the stationary density matrix
within the $S=1$ subspace is $\rho_0=1/3$. Taking the average
over the three eigenvalues of $S_z$ operator, $S_z=0, \pm 1$,
with this density matrix, we see that the dc detector current
in the measurement outcome that corresponds to the $S=1$
subspace is
\begin{equation}
\langle I\rangle = \frac{\Gamma_+ -\Gamma_-}{3}[2 (|t_0|^2+
|\lambda|^2)+|t_0+\lambda|^2+ 8|\delta|^2] ,
\label{ea6} \end{equation}
and can be written as $\langle I\rangle = (I_{\uparrow
\uparrow}+ I_{\downarrow \downarrow}+ I_{\uparrow
\downarrow})/3$.

To calculate the spectral density of the detector current
(\ref{e34}), we consider first {\em non-interacting} qubits
and limit ourselves to the case of weak measurement which can
be described conveniently by going to the basis of the energy
eigenstates. Using the vector intuition for spin operators,
one sees directly that an appropriate rotation of the
Hamiltonian (\ref{e30}) brings it to the form
\begin{equation}
H=-\Omega S_z\, ,\;\;\; \Omega=(\Delta^2+\varepsilon^2)^{1/2}\,,
\label{e32} \end{equation} with three energies $\{-\Omega,
0,\Omega\}$. Upon this rotation, the operator $S_z$ in the
tunneling amplitude (\ref{e31}) changes accordingly
\begin{equation}
S_z \rightarrow (\varepsilon S_z + \Delta S_x)/\Omega \, .
\label{e33} \end{equation}

In the regime of weak measurements, spectral peaks of the detector
output at different frequencies are determined by the evolution
of different groups of matrix elements of the operator $s\equiv
\rho_0 I$, which evolve independently one from another. A peak at
some finite frequency $\bar{\omega}$ is determined by the
off-diagonal matrix elements of $s$ between the states with the
energy difference equal to $\bar{\omega}$. This means that the
total number of the matrix elements relevant for a given
$\bar{\omega}$ coincides with the number of times this interval
occurs in the energy spectrum. In the situation considered in this
subsection, $s=I/3$, where the current operator $I$ is given by
Eq.~(\ref{e34}) in which $S_z$ is transformed according to
Eq.~(\ref{e33}). The structure of the energy levels of the
Hamiltonian (\ref{e32}) implies that there should be three peaks in
the spectrum of the detector output: at $\omega \simeq \Omega,
\, 2\Omega$, and the zero-frequency ``relaxation'' peak.

If the basis states are numbered in the direction of increasing
energy, $H|1\rangle=-\Omega|1\rangle$, $H|2\rangle=0$, etc., the
peak at ${\bf \omega \simeq 2\Omega}$ is determined by the matrix
element $s_{13}$ which satisfy the simple equation that
follows from Eq.~(\ref{e13}):
\begin{equation}
\dot{s}_{13}=(i2\Omega -\gamma) s_{13} \, ,
\label{e35} \end{equation}
and the initial condition $s_{13}(0)=I_{13}/3$, where
\[ I_{13} = a_2 \Delta^2/\Omega^2 \]
is the matrix element of the current operator given by
Eqs.~(\ref{e34}) and (\ref{e33}). The decoherence rate
$\gamma$ in Eq.~(\ref{e35}) is:
\[ \gamma= \frac{1}{2} \big[ \Gamma_{12}+\Gamma_{23}+2
\Gamma_{13} +(\Gamma_+ +\Gamma_-) (4\varepsilon|\delta|/
\Omega)^2\big] \, ,\]
where the last term represents the pure dephasing, and
$\Gamma_{ij}$ is the rate of transitions between the
states $i$ and $j$:
\begin{equation} \begin{array}{ll} \displaystyle
\Gamma_{12} = 2(\Gamma_+ +\Gamma_-) |\delta \Omega+ \lambda
\varepsilon |^2\Delta^2/\Omega^4\, ,\\ \Gamma_{13} =
(\Gamma_+ +\Gamma_-)|\lambda|^2\Delta^4 /\Omega^4 \, ,
\end{array} \label{e36} \end{equation}
and the rate $\Gamma_{23}$ is given by the same expression
as $\Gamma_{12}$ with $\varepsilon \rightarrow -\varepsilon$.
All the rates are obtained from Eq.~(\ref{e13}) in which
the matrix elements of the operator of the the tunnel
amplitude are given by Eqs.~(\ref{e31}) and (\ref{e33}).
Equation (\ref{e35}) means that the spectral peak at $\omega
\simeq 2\Omega$ has simple Lorentzian form:
\begin{equation}
S_I (\omega) = S_0 +\frac{2}{3} \frac{(I_{13})^2 \gamma }{
(\omega-2\Omega)^2 +\gamma^2} \, ,
\label{e37} \end{equation}
and the background noise $S_0$ coincides with Eq.~(\ref{ea6})
for the dc current in which $\Gamma_+ -\Gamma_-$ is replaced
by $\Gamma_+ +\Gamma_-$.

The structure of the ${\bf \omega \simeq \Omega}$ peak of the
spectral density of the detector output is determined by the
time evolution of the matrix elements $s_{12}$ and $s_{23}$
which satisfy the coupled system of equations following from
Eq.~(\ref{e13}):
\begin{equation} \begin{array}{cc} \displaystyle
\dot{s}_{12}=i\Omega s_{12}-\xi_1 s_{12}+ \kappa s_{23}, \\
\dot{s}_{23}=i\Omega s_{23}-\xi_2 s_{23}+\kappa s_{12}.\end{array}
\label{e38} \end{equation}
Initial conditions in these equations are the same as in
Eq.~({\ref{e35}), $s_{ij}(0)=I_{ij}/3$, where the current
matrix elements are:
\[ I_{12} =a_1 (\Delta/\sqrt{2}\Omega) +
a_2 (\sqrt{2}\varepsilon \Delta/\Omega^2) \, , \]
and $I_{23}$ is given by the same expression with $\varepsilon
\rightarrow -\varepsilon$. Again, the relaxation rates $\xi_m$,
$m=1,2$, and the rate $\kappa$ of the ``transfer of coherence''
are obtained by combining the matrix elements of the transmission
amplitude (\ref{e31}), (\ref{e33}) with Eq.~(\ref{e13}):
\begin{eqnarray}
\xi_m=(\Gamma_+ +\Gamma_-) \big[ |\delta |^2(2+
\Delta^2/\Omega^2) - (-1)^m  \nonumber \\
\times 4\mbox{Re}(\delta \lambda^*)  (\varepsilon/\Omega)^3
+ |\lambda|^2 (\Delta^4+\varepsilon^2 \Delta^2+2\varepsilon^4
)/\Omega^4 \big] \, , \nonumber \\
\kappa = (\Gamma_+ +\Gamma_-) ( |\delta |^2 \Omega^2 - |\lambda|^2
\varepsilon^2 \big)(2\Delta^2/\Omega^4) \, . \nonumber
\end{eqnarray} Solving Eq.~(\ref{e38}) by diagonalization of the
matrix of evolution coefficients, we see that the spectral peak at
$\omega \simeq \Omega$ consists in general of two overlapping
Lorentzians with different line-widths and amplitudes:
\begin{equation}
S_I(\omega)=S_0 +\frac{1}{3}\sum_{m=1,2}\frac{A_m \gamma_m
}{(\omega-\Omega)^2 +\gamma_m^2} \, ,
\label{e39} \end{equation}
where
\begin{eqnarray}
&& \gamma_m =(\xi_1+\xi_2)/2 +(-1)^m D\, ,\nonumber \\
&& D\equiv [(\xi_1-\xi_2)^2/4 +\kappa^2]^{1/2} \, , \label{e56} \\
&& A_m= I_1^2+I_2^2-\frac{(-1)^m}{D} [ 2I_1I_2 \kappa
-(I_1^2-I_2^2)\frac{\xi_1-\xi_2}{2} ] \, , \nonumber
\end{eqnarray}
where for later convenience we introduced notations: $I_1\equiv
I_{12}$, $I_2\equiv I_{23}$. In the case of unbiased qubits,
$\varepsilon=0$, the situation becomes much simpler:
$\xi_1=\xi_2$, $I_{12}=I_{23}$, and one of the the Lorentzians
vanishes, $A_2=0$. In this case the peak at $\omega \simeq \Omega$
has the form of one Lorentzian with the line-width $\gamma_1=
(\Gamma_+ +\Gamma_-) (|\delta |^2 + |\lambda|^2)$ and amplitude
$A_1=2a_1^2$. For purely linear measurement: $\lambda=0$,
$|\delta| \ll |t_0|$, this implies that the maximum peak height
$2a_1^2/3\gamma_1$ is limited by $(32/3)S_0$.\cite{lin}

The spectral peak at ${\bf \omega \simeq 0}$ is determined by the
transitions between the energy eigenstates with the rates
(\ref{e36}). The average current $I_{jj}$ in the $j$-th state is
different from the dc current (\ref{ea6}) through the detector:
\begin{eqnarray}
I_{11}-\langle I\rangle &=& a_1(\varepsilon/\Omega) +
a_2 (2\varepsilon^2-\Delta^2)/3\Omega^2 \, , \nonumber \\
I_{22}-\langle I\rangle &=& 2a_2 (\Delta^2-2\varepsilon^2)/
3\Omega^2 \, , \label{ea7} \\
I_{33}-\langle I\rangle &=& - a_1(\varepsilon/\Omega) +
a_2 (2\varepsilon^2-\Delta^2)/3\Omega^2 \, , \nonumber
\end{eqnarray}
Because of nonvanishing
differences $I_{jj}-\langle I\rangle$, transitions between the
eigenstates generate low-frequency noise in the detector current.
Quantitatively, solving the standard kinetic equation for the
evolution of occupation probabilities of the three eigenstates
due to transitions (\ref{e36}) and substituting the solution
into Eq.~(\ref{ea3}) we see that similarly to the peak at
$\omega \simeq \Omega$, the zero-frequency peak of the spectral
density of the detector output consists of two overlapping
Lorentzians with different widths $\eta_m$ and amplitudes $B_m$:
\begin{equation}
S_I(\omega)=S_0 +\frac{1}{3}\sum_{m=1,2}\frac{B_m \eta_m
}{\omega^2 +\eta_m^2} \, ,
\label{e40} \end{equation}
Parameters of the two Lorentians are determined by the transition
rates (\ref{e36}) and the currents (\ref{ea7}):
\begin{eqnarray}
&& \eta_m = \Gamma_{12}+\Gamma_{23}+\Gamma_{13}- (-1)^m D \, ,
\nonumber \\
&& D = [ \Gamma_{12}^2+\Gamma_{23}^2+\Gamma_{13}^2-
\Gamma_{12}\Gamma_{23} - \Gamma_{12} \Gamma_{13}- \Gamma_{13}
\Gamma_{23}]^{1/2} \, , \nonumber \\
&& B_m = \big[c_m a_1^2(\varepsilon/\Omega)^2 + c_{m'}a_2^2
(\Delta^2-2\varepsilon^2)^2/3\Omega^4 \nonumber \\
&& + (-1)^m (\Gamma_{12}- \Gamma_{23}) 2a_1a_2 \varepsilon
(\Delta^2-2\varepsilon^2)/\Omega^3 \big]/D \, , \nonumber \\
&& c_m = 2D +(-1)^m (\Gamma_{12}+\Gamma_{23}-2\Gamma_{13})
\, , \nonumber \end{eqnarray}
where $m'$ is defined as $m'=1,2$, $m'\neq m$.

The situation again simplifies drastically for unbiased qubits,
$\varepsilon \rightarrow 0$, when only one of the Lorentzians has
non-vanishing amplitude $B=2a_2^2/3$. The width of this
non-vanishing Lorentzian is
\[ \eta=3(\Gamma_{12}+ \Gamma_{23})/2 =6(\Gamma_+ +\Gamma_-)
|\delta |^2 \, .\]

Equations (\ref{e37}), (\ref{e39}), and
(\ref{e40}) describe completely the output spectrum of the
detector weakly coupled to identical non-interacting qubits.
In agreement with classical intuition, the spectral peak
at twice the frequency $\Omega$ of the oscillations in
individual qubits appears only if the detector-qubit coupling
is effectively non-linear, when either $\lambda \neq 0$ or
$\delta$ is not too small. The peak at individual qubit
frequency $\Omega$ has generically larger amplitude than the
peak at $2\Omega$ and can be viewed as the result of coherent
superposition of oscillations in two qubits. (Note that the
maximum peak height, $(32/3)S_0$, at $\omega \simeq \Omega$
is larger that two times the height, $4S_0$, of the spectral
peak associated with the oscillations in individual qubits.)
This interpretation is supported by the behavior of the two
peaks as a function of weak {\em qubit-qubit interaction} $\nu$
which we consider now. Including the interaction term in the
Hamiltonian (\ref{e30}) in the evolution equation for the
density matrix in the basis of energy eigenstates (\ref{e32})
one can see that the interaction has no effect on the
dynamics at $\omega\simeq 2\Omega$ until interaction strength
$\nu$ becomes comparable to $\Omega$.

In contrast to this, the dynamics at $\omega\simeq \Omega$ is
affected by much weaker interaction on the order of the
detector-induced dephasing. The interaction breaks coherence
between oscillations in the two qubits and eventually splits the
spectral peak at $\omega \simeq \Omega$ in two as the interaction
strength $\nu$ becomes larger than dephasing. We describe this
splitting quantitatively limiting ourselves to the case of
unbiased qubits, when $\Omega= \Delta$. In the regime of weak
interaction $\nu \ll \Delta$, we can use the ``rotating-wave''
approximation  by keeping only those interaction terms in the
evolution equations that do not mix components oscillating with
different large frequencies on the order of $\Delta$. In this
case, the peak at $\omega \simeq \Delta$ is governed by the
dynamics of the same matrix elements $s_{12}$, $s_{23}$, as for
non-interacting qubits, and Eqs.~(\ref{e38}) for the dynamics of
these elements now are:
\begin{equation} \begin{array}{cc} \displaystyle
\dot{s}_{12}=i(\Delta+\nu/2)s_{12}-\xi s_{12}+ \kappa s_{23},
\\
\dot{s}_{23}=i(\Delta-\nu/2)s_{23}-\xi s_{23}+\kappa s_{12},
\end{array} \label{e41} \end{equation}
where we took into account that for vanishing bias,
\[ \xi_1=\xi_2 =(\Gamma_+ +\Gamma_-)(3|\delta |^2 +
|\lambda|^2) \equiv \xi \, . \]
We see that for $\nu \ll \Delta$, the effect of the interaction
is just the shift of energy of the zero-energy eigenstate of
the Hamiltonian (\ref{e32}) by $\nu/2$ relative to the two
other eigenstates.

\begin{figure}[htb]
\setlength{\unitlength}{1.0in}
\begin{picture}(3.,2.15)
\put(.0,.0){\epsfxsize=2.9in\epsfbox{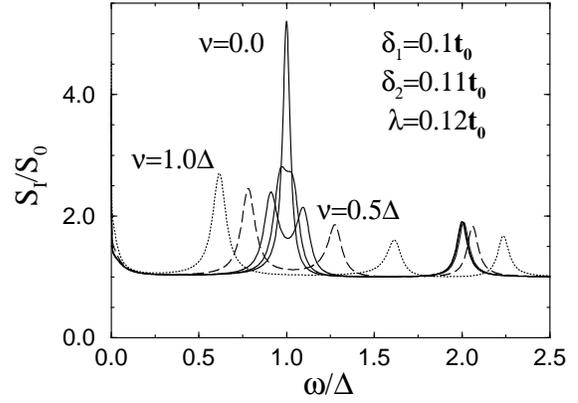}}
\end{picture}
\caption{Evolution of the output spectrum of the non-linear
detector measuring two identical unbiased qubits with the
strength $\nu$ of the qubit-qubit interaction.
The qubit-detector coupling constants $\delta_{1,2}$ are
taken to be slightly different to average the spectrum over
all qubit states. The three solid curves correspond to
$\nu/\Delta =0.0,\,0.1,\,0.2$. In agreement with \protect
Eqs.~(\ref{e42}) -- (\ref{e44}), the peak at $\omega \simeq
\Delta$ is at first suppressed and then split in two by
increasing $\nu$, while the peak at $\omega \simeq 2\Delta$
is not changed noticeably by such a weak interaction.
Dashed and dotted lines show the regime of relatively
strong interaction: $\nu/\Delta =0.5$ and $\nu/\Delta=1.0$,
respectively, that is described by Eqs.~(\ref{e46}) and
(\ref{e47}).}
\label{fig5} \end{figure}

Equations (\ref{e41}) have the same structure as Eqs.~(\ref{e38}),
and for weak interaction, $\nu/2< \kappa= 2 (\Gamma_+ +\Gamma_-)
|\delta |^2$, result in the same form of the peak of the spectral
density at $\omega \simeq \Delta$ as in Eq.~(\ref{e39}): two
overlapping Lorentzians with different line-width $\gamma_m$ and
amplitudes $A_m$, $m=1,2$:
\begin{equation} \begin{array}{cc} \displaystyle
\gamma_m =\xi +(-1)^m D\, , \;\;\; D= [\kappa^2-\nu^2/4]^{1/2}
\, ,  \\ A_m= a_1^2 (1-(-1)^m\kappa/D) \, .
\end{array} \label{e42} \end{equation}
If $\nu/2\rightarrow \kappa$, then $D\rightarrow 0$, and the two
Lorentzians combine to give the peak of the following form:
\begin{equation}
S_I(\omega)=S_0 +\frac{2a_1^2}{3}(1-\kappa \frac{\partial
}{\partial \xi}) \frac{\xi}{(\omega-\Delta)^2 +\xi^2} \, .
\label{e43} \end{equation} For stronger interaction, $\nu/2>
\kappa$, the two Lorentzians acquire different frequencies $\Delta
\pm D$, $D= [\nu^2/4 - \kappa^2]^{1/2}$, while having the same
line-width $\xi$, so that the total spectral density at $\omega
\simeq \Delta$  is:
\begin{equation}
S_I(\omega)=S_0 +\frac{a_1^2}{3}\sum_{\pm}\frac{\xi +\kappa
\pm (\Delta-\omega) \kappa/D }{(\omega-\Delta \mp D)^2 +\xi^2}
\, .
\label{e44} \end{equation}
For $\nu \gg \kappa$, spectral density (\ref{e44}) consists of
two independent Lorentzians separated by the large frequency
interval $\nu$. The amplitude of both peaks is $a_1^2/3\xi$
and is limited by $(16/9) S_0$, the value that should be
reached by an ideal linear detector.

Equations (\ref{e42}), (\ref{e43}), and (\ref{e44}) are valid in
the regime of weak interaction, $\nu \simeq \kappa \ll \Delta$,
when interaction affects strongly only the spectral peak at
$\omega\simeq \Delta$, which is first broadened and then split in
two with increasing interaction strength. To describe the detector
output spectrum for stronger interaction $\nu \simeq \Delta \gg
\kappa$, we need to diagonalize the Hamiltonian (\ref{e30}) with
non-vanishing $\nu$. This can be done easily in the case of
unbiased qubits $\varepsilon =0$, when the eigenstates of the
Hamiltonian (\ref{e30}) coincide with the two eigenstates of the
Hamiltonian $H_+$ (\ref{e17}) and the state $(|\!\uparrow \uparrow
\rangle - |\!\downarrow \downarrow \rangle)/\sqrt{2}$ of the
subspace $D_-$ introduced before in Eq.~(\ref{e16}). The
eigenenergies of these states are:
\begin{equation}
\{ -\Omega_+/2\, , \; \nu/2\, , \; \Omega_+/2 \, \} , \;\;\;\;
\Omega_+= (4\Delta^2 +\nu^2)^{1/2} \, .
\label{e45} \end{equation}

The three finite-frequency peaks in the spectrum correspond to
three energy intervals, $\Omega_+$, $(\Omega_+ \pm \nu)/2$ in the
spectrum of these eigenstates. Since all energy intervals are
different and therefore there is no transfer of coherence between
them that exists, e.g., in Eqs.~(\ref{e38}), the peaks are simple
Lorentzians. Transforming the tunneling amplitude (\ref{e31}) into
the basis of eigenstates (\ref{e45}), we find all the rates in the
evolution Eq.~(\ref{e13}) and the matrix elements of the current
operator (\ref{e34}) in this basis and find the parameters of
these Lorentzians: amplitude $A$ and the line-width $\gamma$, both
defined as in Eq.~(\ref{e39}):
\begin{eqnarray}
&& \omega \simeq  \Omega_+\, , \;\;\;\; A= 8(a_2 \Delta/\Omega_+
)^2 \, , \nonumber \\
\; \gamma &=& (\Gamma_+ +\Gamma_-) [2|\delta|^2+(1+\nu^2/\Omega_+^2)
|\lambda|^2 ] \, ,  \label{e46} \\
&& \omega \simeq  (\Omega_+ \pm \nu)/2 \, , \;\;\;\; A=a_1^2
(1\mp\nu/\Omega_+) \, , \nonumber \\
\; \gamma &=& (\Gamma_+ +\Gamma_-) [3(1\mp \nu/\Omega_+)|\delta|^2+
(1\pm \nu/\Omega_+) |\lambda|^2 ] \, . \label{e47}
\end{eqnarray}
One can see that parameters of the two lower-frequency peaks
(\ref{e47}) agree in the regime $\Delta \gg \nu \gg \kappa$ with
the peaks in the spectral density (\ref{e44}), while the peak
(\ref{e46}) at the largest frequency $\Omega_+$ coincides with that
described by Eq.~(\ref{e37}) the case of unbiased non-interacting
qubits.

Figure \ref{fig5} illustrates evolution of the output spectrum
of the non-linear detector measuring identical qubits due to
changing interaction strength. We see that
this evolution agrees with the analytical description developed
above. Weak qubit-qubit interaction $\nu \simeq \kappa \ll \Delta$
suppresses and subsequently splits the spectral peak at
$\omega \simeq \Omega$ while not changing the peak $\omega \simeq
2\Omega$. Stronger qubit-qubit interaction $\nu \simeq \Delta \gg
\kappa$ shifts the $\omega \simeq 2\Omega$ peak to higher
frequencies while moving the two peaks around $\omega \simeq
\Omega$ further apart.

\subsection{Non-linear detector measuring different qubits}

As the last example of the output spectrum of the non-linear
detector measuring two qubits we consider the most general
situation when both the tunneling amplitudes and the
detector-qubit coupling constants are different for the two
qubits. We assume that the qubit parameters are such that all
energy intervals in the spectrum of eigenstates are larger than
the back-action dephasing rate, i.e., the detector-qubit coupling
is weak. We begin with the case of {\em unbiased qubits},
$\varepsilon_j=0$, when the coherent oscillations in the two
qubits should have the largest amplitude. At this bias point, the
Hamiltonian (\ref{e2}) breaks into two subspaces $D_{\pm}$
(\ref{e16}) and can be diagonalized directly. It is convenient to
order the eigenstates taking into account these subspaces:
\begin{equation}
\{ \Omega_+/2\, , \; -\Omega_+/2\, , \; \Omega_-/2\, ,
-\Omega_-/2 \, \} \, .
\label{e50} \end{equation}
The energies $\Omega_{\pm}$ are defined in Eq.~(\ref{e49}). The
wavefunctions of the eigenstates numbered in this order are:
\begin{eqnarray*}
\;\; |\psi_1\rangle = [(\alpha +\beta) (|\! \uparrow
\uparrow \rangle +|\! \downarrow \downarrow \rangle) +
(\beta -\alpha) (|\! \uparrow \downarrow \rangle
+ |\! \downarrow \uparrow \rangle) ]/2\, , \\
\;\; |\psi_2\rangle = [(\alpha -\beta) (|\! \uparrow
\uparrow \rangle +|\! \downarrow \downarrow \rangle) +
(\alpha +\beta) (|\! \uparrow \downarrow \rangle +
|\! \downarrow \uparrow \rangle ) ]/2 \, ,  \\
\;\; |\psi_3\rangle =[(\bar{\alpha} +\bar{\beta}) (|\! \uparrow
\uparrow \rangle -|\! \downarrow \downarrow \rangle ) +
(\bar{\beta} -\bar{\alpha} )
(|\!\downarrow \uparrow \rangle - |\! \uparrow \downarrow
\rangle )]/2\, ,  \\
|\psi_4\rangle =[(\bar{\alpha} -\bar{\beta}) (|\! \uparrow
\uparrow \rangle -|\! \downarrow \downarrow \rangle ) +
(\bar{\alpha} +\bar{\beta})
(|\! \downarrow \uparrow \rangle - |\! \uparrow \downarrow
\rangle) ]/2 \, ,
\end{eqnarray*}
where
\begin{eqnarray*}
\alpha, \beta = (1/\sqrt{2}) [1\pm (\Delta_1+ \Delta_2)/\Omega_+
]^{1/2}\, ,\\
\bar{\alpha}, \bar{\beta} = (1/\sqrt{2}) [1\pm (\Delta_1-\Delta_2)/
\Omega_-]^{1/2} \, .
\end{eqnarray*}

Assuming that there are no ``accidental'' degeneracies, there
are four different energy intervals in the spectrum (\ref{e50})
which should be reflected as four finite-frequency peaks in the
detector output spectrum. For different qubit parameters, there
are no selection rules and the detector back-action mixes all four
state of the qubit system. The stationary density matrix is then
given by Eq.~(\ref{e14}) and the background detector noise $S_0$
corresponds to the dc current (\ref{e15}). The largest
($\Omega_+$) and the smallest ($\Omega_-$) energy intervals occur
only once in the spectrum (\ref{e50}), so that the peaks at
$\omega \simeq \Omega_{\pm}$ has the shape of simple Lorentians:
\begin{equation}
\omega \simeq \Omega_{\pm} \, , \;\;\; S_I(\omega)=S_0 +
\frac{1}{4}\frac{A_{\pm} \gamma_{\pm} }{(\omega-\Omega_{\pm})^2
+\gamma_{\pm}^2} \, .
\label{e51} \end{equation}

Calculating the current matrix elements and all the rates in
Eq.~(\ref{e13}) for evolution of the density matrix in the basis
of eigenstates (\ref{e50}) similarly to what is done for the
description of decoherence in coupled qubits \cite{dec}, we find
parameters of the Lorentzians in Eq.~(\ref{e51}). It is convenient
to express the peak amplitudes $A_{\pm}$ through the characteristic
amplitudes of current modulation by the qubits analogous to the
amplitudes $a_{1,2}$ (\ref{e48}) for identical qubits. In the case
of different strength of the detector-qubit coupling for the two
qubits, we have three such amplitudes:
\begin{eqnarray}
a_{11} &=& (I_{\uparrow \uparrow}+I_{\uparrow \downarrow}-
I_{\downarrow \uparrow }-I_{\downarrow \downarrow})/4 \nonumber \\
&=& 2(\Gamma_+ -\Gamma_-) \mbox{Re} [t_0\delta_1^* + \lambda
\delta_2^*]\, , \nonumber \\
a_{12} &=& (I_{\uparrow \uparrow}+I_{\downarrow \uparrow }-
I_{\uparrow \downarrow}- I_{\downarrow \downarrow})/4 \label{e52} \\
&=& 2(\Gamma_+ -\Gamma_-) \mbox{Re} [t_0\delta_2^*+\lambda
\delta_1^*]\, , \nonumber \\
a_2 &=& (I_{\uparrow \uparrow}+I_{\downarrow \downarrow}-
I_{\uparrow \downarrow}- I_{\downarrow \uparrow })/4 \nonumber \\
&=& 2(\Gamma_+ -\Gamma_-) \mbox{Re} [t_0\lambda^* +\delta_1
\delta_2^*] \, . \nonumber
\end{eqnarray}
For identical qubits, $a_{11}$ and $a_{12}$ reduce to $a_1/2$,
while the definition of $a_2$ in Eqs.~(\ref{e52}) and (\ref{e48})
coincide. Qualitatively, $a_{1j}$ corresponds to the amplitude of
modulation of the detector current due to oscillations in the
$j$-th qubit, and $a_2$ - is the similar amplitude due to
``collective'' oscillations of the two qubits between parallel and
anti-parallel configurations. In terms of the amplitudes
(\ref{e52}), $A_{\pm}$ are given by simple expressions:
\begin{equation}
A_{\pm}= 2a_2^2 [(\Delta_1\pm \Delta_2)/\Omega_{\pm}]^2
\label{e53} \end{equation} The line-widths $\gamma_{\pm}$ in
Eq.~(\ref{e51}) are:
\begin{equation}
\gamma_{\pm}= (\Gamma_+ +\Gamma_-) [|\delta_1|^2+|\delta_2|^2+
|\lambda|^2(1+ \nu^2/\Omega_{\pm}^2)] \, .
\label{e54} \end{equation}

In contrast to the energy intervals $\Omega_{\pm}$ which occur
only once in the energy spectrum (\ref{e50}), the intervals
$(\Omega_+ \pm\Omega_-)/2$ occur twice each. This means that
the spectral peaks at $\omega \simeq (\Omega_+ \pm\Omega_-)/2$
are not simple Lorentzians, and their shape is controlled
by the two matrix elements of the density matrix evolving
according to a system of coupled equations identical with
Eqs.~(\ref{e41}). This means that these peaks consist of two
overlapping Lorentzians each, and the output spectrum in their
vicinity is given by the equation similar to Eq.~(\ref{e39}):
\[ S_I(\omega) =S_0 +\frac{1}{4}\sum_{m=1,2}\frac{A_m^{(\pm)}
\gamma_m^{(\pm)} }{[\omega-(\Omega_+ \pm\Omega_-)/2]^2 +
[\gamma_m^{(\pm)}]^2} \, , \] for $\omega \simeq (\Omega_+
\pm\Omega_-)/2$. Here the amplitudes $A_m^{(\pm)}$ and line-widths
$\gamma_m^{(\pm)}$ are given by the same Eqs.~(\ref{e56}), where
now
\begin{eqnarray}
\xi_m^{(\pm)}=(\Gamma_+ +\Gamma_-) \big[ (1\pm \frac{\nu^2}{
\Omega_+\Omega_-})|\lambda|^2+|\delta_1 |^2+|\delta_2 |^2
\nonumber \\ - (-1)^m (\frac{\nu}{\Omega_+} \mp \frac{\nu}{
\Omega_-}) \mbox{Re} (\delta_1 \delta_2^*)\big] \nonumber \\
\kappa^{(\pm)} =\frac{1}{2} (\Gamma_+ +\Gamma_-) \big[\frac{
\Delta_1^2-\Delta_2^2}{\Omega_+ \Omega_-}(|\delta_1 |^2+
|\delta_2 |^2) \label{e57} \\
-(\frac{\nu^2}{\Omega_+\Omega_-} \pm 1)(|\delta_1 |^2-
|\delta_2 |^2) \big] \, , \nonumber \\
I_m^{(\pm)}= (\pm 1)^{m+1} \frac{a_{11}}{\sqrt{2}}\big[1 \mp
\frac{\nu^2-\Delta_1^2+\Delta_2^2}{\Omega_+\Omega_-}
\big]^{1/2} \nonumber \\
+ (\mp 1)^m\frac{a_{12}}{\sqrt{2}}\big[1 \mp \frac{\nu^2+
\Delta_1^2-\Delta_2^2}{\Omega_+\Omega_-} \big]^{1/2} \, .
\nonumber  \end{eqnarray}

In the case of non-interacting qubits and no quadratic coupling,
$\nu=\lambda=0$, the peaks at $(\Omega_+ \pm\Omega_-)/2$ represent
oscillations in the individual qubits: the peak at $(\Omega_+ -
\Omega_-)/2$ describes oscillations in a qubit with a smaller
$\Delta$, while the peak at $(\Omega_+ +\Omega_-)/2$ -- in the
qubit with larger $\Delta$. Only one Lorentzian ($m=1$) is then
non-vanishing for each peak, and has the same parameters as it
would have in the absence of the other qubit. For instance, if
$\Delta_1>\Delta_2$, we get from Eqs.~(\ref{e56}) and (\ref{e57})
for $\nu=\lambda=0$:
\begin{eqnarray*}
(\Omega_+ +\Omega_-)/2=\Delta_1, \; \gamma^{(+)}_1 =
(\Gamma_+ +\Gamma_-)|\delta_1 |^2, \; A^{(+)}_1= 4a_{11}^2, \\
(\Omega_+ -\Omega_-)/2=\Delta_2, \; \gamma^{(-)}_1 =
(\Gamma_+ +\Gamma_-)|\delta_2 |^2, \; A^{(-)}_1= 4a_{12}^2.
\end{eqnarray*}
This behavior is the natural consequence of the linearity of the
detector-qubit system for $\nu=\lambda=0$. Non-vanishing
qubit-qubit interaction and/or non-linear detector response modify
parameters of these ``linear'' peaks according to Eqs.~(\ref{e56})
and (\ref{e57}). In contrast to these peaks, the spectral peaks at
$\Omega_{\pm}$ reflect directly the detector non-linearity and
vanish in the linear regime when $\lambda=0$ and $\delta_j \ll
t_0$. As one can see from Eq.~(\ref{e52}), the characteristic
amplitude $a_2$ of these peaks scales for $\lambda=0$ as a higher
power of $\delta_j$ than the amplitudes $a_{1j}$ of the linear
peaks.

An example of the output spectrum of the non-linear detector
measuring unbiased qubits with different tunneling amplitudes is
shown in Fig.~\ref{fig6}. One can see that when the linear and
non-linear coefficient of the detector-qubit coupling are roughly
similar, the linear peaks are more pronounced than the peaks at
combination frequencies. Qubit-qubit interaction shifts all but
the lower-frequency linear peak up in frequency and reduces
both the amplitudes of the higher-frequency peaks and the
distance between them.

\begin{figure}[htb]
\setlength{\unitlength}{1.0in}
\begin{picture}(3.,2.)
\put(.1,.0){\epsfxsize=2.7in\epsfbox{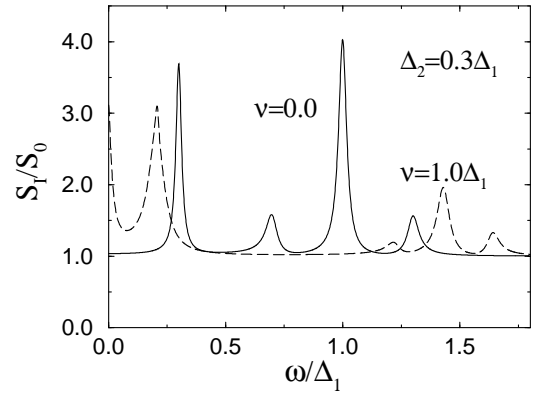}}
\end{picture}
\caption{Output spectra of the non-linear detector measuring
two different unbiased qubits. Solid line is the spectrum in
the case of non-interacting qubits. The two larger peaks are
the ``linear'' peaks that correspond to the oscillations in
the individual qubits, while smaller peaks are non-linear
peaks at the combination frequencies. Dashed line is the
spectrum for interacting qubits. Interaction shifts the
lower-frequency liner peak down and all other peaks up in
frequency. Parameters of the detector-qubit coupling are:
$\delta_1=0.12t_0,\, \delta_2=0.09t_0,\, \lambda=0.08t_0$.}
\label{fig6} \end{figure}

Analytical results of this subsection can be easily extended to
the finite qubit bias if the qubits are {\em non-interacting}.
The eigenstates of the Hamiltonian (\ref{e2}) are then the
products of the eigenstates of the individual qubits, so that
the energy spectrum of the two-qubit system is:
\begin{equation}
(1/2)\{ \Omega_1+\Omega_2, \, \Omega_1-\Omega_2, \,
\Omega_2-\Omega_1, \, -(\Omega_1+\Omega_2) \, \} \, , \label{e60}
\end{equation} where $\pm \Omega_j/2$ with
$\Omega_j=(\varepsilon_j^2+ \Delta_j^2)^{1/2}$ are the
eigenenergies of the $j$-th qubit.

Similarly to the spectrum (\ref{e50}), there are four different
energy intervals in (\ref{e60}). Two of them, $\Omega_1\pm \Omega_2
\equiv \Omega_{\pm}$ occur only once and are reflected as simple
Lorentzian peaks in the detector output spectrum. These
``non-linear'' peaks represent the mixture of the individual qubit
oscillations with frequencies $\Omega_{1,2}$ and have non-vanishing
amplitude only if the detector response is non-linear.
Quantitatively, the output spectrum around this peaks can be
written as in Eq.~(\ref{e51}), where now the peak parameters are:
\begin{eqnarray}
A_{\pm}= 2a_2^2 \big(\frac{ \Delta_1\Delta_2}{\Omega_1\Omega_2}
\big)^2 \, , \;\;\;\; \gamma_{\pm}= (\Gamma_+ +\Gamma_-)
\Big[ |\lambda|^2 \cdot \nonumber \\
\big[1- \big(\frac{\varepsilon_1\varepsilon_2}{ \Omega_1
\Omega_2} \big)^2 \big] + |\delta_1|^2 + |\delta_2|^2 +\big|
\delta_1 \frac{\varepsilon_1}{\Omega_1} \pm \delta_2
\frac{\varepsilon_2 }{\Omega_2}\big|^2 \Big] \, .
\label{e61} \end{eqnarray}
They are found by calculating both the rates in Eq.~(\ref{e13})
for the density matrix, and the matrix elements of the current
operator $I$:
\[I= \langle I\rangle +\sum_j a_{1j} \bar{\sigma}_z^j + a_2
\bar{\sigma}_z^1\bar{\sigma}_z^2 \, , \;\;\; \bar{\sigma}_z^j=
\frac{1}{\Omega_j} (\varepsilon_j\sigma_z^j + \Delta_j
\sigma_x^j)\, ,\]
in the basis of states (\ref{e60}). In this expression,
$\langle I\rangle$ is the average current (\ref{e15}).

The energy intervals $\Omega_j$ occur twice each in the spectrum
(\ref{e60}). This means that the spectral peaks at $\omega \simeq
\Omega_j$ that correspond to the individual qubit oscillations
have the same form as in Eq.~(\ref{e39}), so that
\begin{equation}
S_I(\omega) =S_0 +\frac{1}{4}\sum_{m=1,2}\frac{A_{jm}\gamma_{jm}
}{(\omega-\Omega_j )^2 + \gamma_{jm}^2} \label{e62} \end{equation}
for $\omega \simeq \Omega_j$, where the amplitudes $A_{jm}$ and
line-widths $\gamma_{jm}$ are given by Eqs.~(\ref{e56}) with
\begin{eqnarray}
\xi_{jm}=(\Gamma_+ +\Gamma_-) \big[ (1+[\varepsilon_j/\Omega_j]^2 )
\big|\delta_j -(-1)^m\lambda (\varepsilon_{j'}/\Omega_{j'}) \big|^2
\nonumber \\
+ (|\lambda|^2+|\delta_{j'} |^2)(\Delta_{j'}/\Omega_{j'})^2 \big]
\, , \nonumber \\
\kappa_j =(\Gamma_+ +\Gamma_-) (\Delta_{j'}/\Omega_{j'})^2
[|\delta_{j'} |^2+ |\lambda |^2 (\varepsilon_j/\Omega_j)^2 ]\, ,
\nonumber \\
I_{jm}= \big[a_{1j}-(-1)^m a_2 (\varepsilon_{j'}/\Omega_{j'})
\big](\Delta_j/\Omega_j) \, ,
\nonumber  \end{eqnarray}
where $j'\neq j$.

We see from Eq.~(\ref{e62}) that for non-interacting qubits, the
non-vanishing qubit bias just shifts the frequency position of the
liner peaks (\ref{e57}) without qualitatively changing their
shape. If both the bias and the qubit-qubit interaction are
finite, the bias splits each of the linear peaks in two simple
Lorentzians bringing the total number of the finite-frequency
peaks in the spectrum of the detector output to six as it should
be in the generic situation (see, e.g., Fig.~\ref{fig3}).

\section{Conclusion}

To summarize, we have developed a theory of continuous quantum
measurements of coherent oscillations in two coupled qubits by a
non-linear detector. Calculated spectra of the detector output
show that the detector non-linearity leads to the appearance of
the spectral peaks at the combination frequencies of the qubit
oscillations in the detector output. The spectra have the
non-trivial dependence on the strength of the qubit-qubit
interaction. For identical non-interacting qubits, the spectral
peaks at frequency of individual qubit oscillations are
superimposed coherently, with weak interaction breaking this
coherent superposition and splitting the oscillation peak. In
general, qubit-qubit interaction should manifest itself
qualitatively through the total number of peaks in the output
spectrum: the total number of peaks is at most 4 in the case of
non-interacting qubits, while it can reach 6 in the most general
situation with both non-vanishing qubit-qubit interaction and the
detector non-linearity.

\section{Acknowledgments}
W.M. and D.V.A. would like to thank A.N. Korotkov and R. Ruskov
for useful discussions at the initial stages of this work. This
work was supported in part by the NSF under grant \# 0121428 and
by ARDA and DOD under the DURINT grant \# F49620-01-1-0439 (W.M.
and D.V.A.), and by the EC contracts IST-SQUBIT2 and RTN-Nanoscale
Dynamics (F.P. and R.F.).


\begin{references}
\bibitem{rem2} For measurements of other systems (different from
the two qubits) expansion in the measured operators similar to
Eq.~(\ref{e4}) can be justified as the Taylor's expansion in weak
detector-system coupling. Higher-order terms in this expansion
should manifest themselves in measurements of larger number of
qubits.
\end{references}
\end{document}